\documentclass[12 pt, amsfonts, amssymb,color]{article}

\usepackage{amsmath,amssymb,amsfonts,latexsym,float,graphics,epsfig}
\usepackage{subfig}
\usepackage{verbatim}

\begin{document}

\renewcommand\theequation{\arabic{section}.\arabic{equation}}
\catcode`@=11 \@addtoreset{equation}{section}
\newtheorem{axiom}{Definition}[section]
\newtheorem{theorem}{Theorem}[section]
\newtheorem{axiom2}{Example}[section]
\newtheorem{lem}{Lemma}[section]
\newtheorem{prop}{Proposition}[section]
\newtheorem{cor}{Corollary}[section]
\newcommand{\be}{\begin{equation}}
\newcommand{\ee}{\end{equation}}

\newcommand{\equal}{\!\!\!&=&\!\!\!}
\newcommand{\rd}{\partial}
\newcommand{\g}{\hat {\cal G}}
\newcommand{\bo}{\bigodot}
\newcommand{\res}{\mathop{\mbox{\rm res}}}
\newcommand{\diag}{\mathop{\mbox{\rm diag}}}
\newcommand{\Tr}{\mathop{\mbox{\rm Tr}}}
\newcommand{\const}{\mbox{\rm const.}\;}
\newcommand{\cA}{{\cal A}}
\newcommand{\bA}{{\bf A}}
\newcommand{\Abar}{{\bar{A}}}
\newcommand{\cAbar}{{\bar{\cA}}}
\newcommand{\bAbar}{{\bar{\bA}}}
\newcommand{\cB}{{\cal B}}
\newcommand{\bB}{{\bf B}}
\newcommand{\Bbar}{{\bar{B}}}
\newcommand{\cBbar}{{\bar{\cB}}}
\newcommand{\bBbar}{{\bar{\bB}}}
\newcommand{\bC}{{\bf C}}
\newcommand{\cbar}{{\bar{c}}}
\newcommand{\Cbar}{{\bar{C}}}
\newcommand{\Hbar}{{\bar{H}}}
\newcommand{\bL}{{\bf L}}
\newcommand{\Lbar}{{\bar{L}}}
\newcommand{\cLbar}{{\bar{\cL}}}
\newcommand{\bLbar}{{\bar{\bL}}}
\newcommand{\cM}{{\cal M}}
\newcommand{\bM}{{\bf M}}
\newcommand{\Mbar}{{\bar{M}}}
\newcommand{\cMbar}{{\bar{\cM}}}
\newcommand{\bMbar}{{\bar{\bM}}}
\newcommand{\cP}{{\cal P}}
\newcommand{\cQ}{{\cal Q}}
\newcommand{\bU}{{\bf U}}
\newcommand{\bR}{{\bf R}}
\newcommand{\cW}{{\cal W}}
\newcommand{\bW}{{\bf W}}
\newcommand{\bZ}{{\bf Z}}
\newcommand{\Wbar}{{\bar{W}}}
\newcommand{\Xbar}{{\bar{X}}}
\newcommand{\cWbar}{{\bar{\cW}}}
\newcommand{\bWbar}{{\bar{\bW}}}
\newcommand{\abar}{{\bar{a}}}
\newcommand{\nbar}{{\bar{n}}}
\newcommand{\pbar}{{\bar{p}}}
\newcommand{\tbar}{{\bar{t}}}
\newcommand{\ubar}{{\bar{u}}}
\newcommand{\utilde}{\tilde{u}}
\newcommand{\vbar}{{\bar{v}}}
\newcommand{\wbar}{{\bar{w}}}
\newcommand{\phibar}{{\bar{\phi}}}
\newcommand{\Psibar}{{\bar{\Psi}}}
\newcommand{\bLambda}{{\bf \Lambda}}
\newcommand{\bDelta}{{\bf \Delta}}
\newcommand{\p}{\partial}
\newcommand{\om}{{\Omega \cal G}}
\newcommand{\ID}{{\mathbb{D}}}
\newcommand{\pr}{{\prime}}
\newcommand{\prr}{{\prime\prime}}
\newcommand{\prrr}{{\prime\prime\prime}}
\title{ Cheillini integrability and quadratically damped oscillators}
\author{Ankan Pandey$^1$\footnote{E-mail: ankan.pandey@bose.res.in}, 
A Ghose Choudhury$^2$\footnote{E-mail aghosechoudhury@gmail.com}, 
Partha Guha$^{1}$\footnote{E-mail: partha@bose.res.in}\\
\\
$^1$ S N Bose National Centre for Basic Sciences\\ 
JD Block, Sector III, Salt Lake, Kolkata 700106,  India\\
$^2$ Department of Physics, Surendranath  College\\ 24/2 Mahatma
Gandhi Road, Calcutta 700009, India\\}

\date{ }

 \maketitle

\smallskip

\smallskip

\begin{abstract}
\textit{ In this paper a new approach to study an equation of the Li\'enard type with a strong quadratic damping 
is proposed based on Jacobi's last multiplier and Cheillini's integrability condition. We obtain a closed form solution
of the transcedental characteristic equation of the Li\'enard type equation using the Lambert $W$-function.}
\end{abstract}

\smallskip

\paragraph{Mathematics Classification (2010)}:34C15, 34C20.

\smallskip


\section{Introduction}

In recent times a number of articles have appeared in the literature which deal with the phenomenon of a linear 
oscillator subject to a quadratic damping force \cite{Cveticanin1,Cveticanin2,Fay,Klotter,KR}. Most elementary 
textbooks deal 
with viscous damping  for the obvious reason that it involves a linear dependance on the velocity of the 
oscillator and presents the simplest situation where an exact analytical treatment is possible. In general this 
involves analysis of a second-order ordinary 
differential equation (ODE) of the Li\'{e}nard type \cite{Lienard}, namely $\ddot{x}+f(x)\dot{x}+g(x)=0$, where 
it is assumed that $f$ is a constant and the function $g=x$.  As damping does not arise from a single physical 
phenomena and is itself of various kinds, e.g., material damping, structural damping, interfacial damping, 
aerodynamic and hydrological drag etc., therefore a different mathematical description
is needed in each case. Systems like the simple harmonic oscillator and the viscously damped harmonic oscillator, 
both of which can be solved by standard undergraduate mathematical techniques, however, represent idealizations of 
real life phenomena because they ignore nonlinear aspects of the forcing term as well as the damping force. 
A more realistic model in which damping is proportional to the square of the velocity is usually common at 
higher velocities and is applicable to problems involving hydrological drag and in aerodynamics. 
When an immersed object moves through a fluid at relatively high Reynolds numbers \cite{NM}- 
the corresponding drag force is found to be proportional to the square of the velocity $v = sgn( \dot{x})\dot{x}^2$. 
Oscillators with a non-negative real-power
restoring force $F(x) = k sgn(x)|x|^{\alpha}$ and quadratic damping have been recently studied by Kovacic and Rakaric \cite{KR}.

\smallskip

The principal feature  associated with quadratic damping is a discontinuous jump of the damping force in the 
equation of motion whenever the velocity vanishes such that the frictional force always opposes the motion. 
In case of oscillatory systems this occurs every half cycles and means that instead of a single equation of 
motion the latter splits into two parts depending on the sign of the velocity. Each equation has to solved 
separately and matched at the points where the velocity changes sign. In general solving such a system in 
presence of nonlinearity proves to be a rather daunting task and only in rare cases is an exact solution 
to be expected. Numerical techniques on the other hand provide valuable information  about the 
evolution of the system and its general nature. \\

From the mathematical point of view the construction of first integrals for systems involving a quadratic 
dependance on the velocity often provides interesting insights. Indeed constants of motion are the bed 
rock of many of the conservation principles at the heart of theoretical physics: the work-energy 
theorem applied to a conservative system, is perhaps the most striking and oft quoted example, 
as it has evolved into the principle of conservation of energy.\\

In this paper we examine the equation , $\ddot{x}+sgn(\dot{x}) f(x) \dot{x}^2 +g(x)=0$, 
in the light of several recent articles which have also dealt 
with the same equation. This is a discontinuous generalization of an equation of the Li\'enard type 
involving a quadratic dependence on the velocity. The issue of isochronicity
in such equations was extensively studied in \cite{Sabatini,CG,GC}.
In particular we show that by imposing the Cheillini condition of integrability on the 
functions $f$ and $g$ one can subsume many of the previous examples into a compact scheme. 
Incidentally the Cheillini condition is typically encountered in the context of 
integrability of the standard Li\'{e}nard equation in course of its transformation to the 
first-order Abel equation of the first kind and also while finding a Lagrangian/Hamiltonian 
description of the Li\'{e}nard equation \cite{Harko2}. However, its application to the case of quadratic 
damping appears to be new. We show how one can derive in a systematic manner the maximum 
amplitudes analytically in terms of the Lambert $W$ function.\\

The organization of the paper is as follows. In Section 2 we review a second-order ODE with a 
quadratic dependence on velocity in the context of its Lagrangian/Hamiltonian description. 
It is shown that such a  system may be interpreted as one displaying a position dependent mass function. 
The trajectory is explicitly displayed by numerical investigations. 
In Section 3 we split the ODE into two parts as mentioned above depending on the sign of the 
velocity $\dot{x}$ and investigate the trajectories, maximum amplitudes as well as  period of oscillations. 
In particular we show that the periods of the cycles and the corresponding maximum amplitudes are both 
determined exclusively by a potential function which involves the position dependent mass function. 
Furthermore by invoking the Cheillini integrability condition it is possible to write down analytic 
formulae for the maximum amplitudes in terms of the Lambert $W$ function \cite{Corless},
named after the eighteenth century scientist J. H. Lambert \cite{Lambert}. 
The Lambert $W$ function is defined as the inverse function of the $x \mapsto xe^x$ mapping and thus solves the equation.
$ye^y = x$ equation. This solution is given in the 
form of the Lambert $W$ function, $y = W(x)$, i.e. $W$ satisfies $W(x)e^{W(x)} = x$. 
This equation always has an infinite number of solutions, most of which are
complex, and $W$ is a multivalued function. The examples presented here also includes those obtained 
earlier by Cveticanin \cite{Cveticanin1,Cveticanin2}.

\section{The Hamiltonian in presence of quadratic velocity}

Consider a second-order ODE with a quadratic dependance on the velocity given by
\be\label{E1} \ddot{x}+f(x)\dot{x}^2 +g(x)=0.\ee
we assume $f(x) $ and $g(x)$ are such that $f(0)=g(0)=0$ and $f(x)$ is integrable while $g^\prime(0)>0$. The functional form of $g(x)=g^\prime(0) x +g_n(x)$ where $g_n(x)$ is analytic. As demonstrated in \cite{NL3,NUT}, the Jacobi Last Multiplier (JLM) provides a convenient tool for obtaining a Lagrangian for second-order equations of the form $\ddot{x}=\mathcal{F}(x, \dot{x})$. It is defined as a solution of
\be\label{E2} \frac{d}{dt} \log M+\frac{\partial \mathcal{F}(x, \dot{x})}{\partial \dot{x}}=0.\ee
In the present case it follows that
\be\label{E3} M=\exp\left(2F(x)\right), \;\;\;\; \hbox{ where }  \;\;\;\;  F(x)=\int_0^x f(s) ds.\ee
The relationship between the JLM, $M$ and the Lagrangian is provided by $M=\partial^2 L/\partial \dot{x}^2$ as a consequence of which for (\ref{E1})  the Lagrangian may be expressed in the form
\be\label{E4} L=\frac{1}{2}e^{2F(x)} \dot{x}^2 -V(x),\ee
where $V(x)$ is determined by substituting (\ref{E4}) into the Euler-Lagrange equation $$\frac{d}{dt}\left(\frac{\partial L}{\partial \dot{x}}\right)=\left(\frac{\partial L}{\partial x}\right),$$ which immediately gives
\be\label{E5} V(x)=\int_0^x e^{2F(s)} g(s) ds.\ee
By means of the standard Legendre transformation we can obtain the Hamiltonian as
\be\label{E6}H=\frac{1}{2}e^{2F(x)} \dot{x}^2+\int_0^x e^{2F(s)} g(s) ds.\ee It is easily verified that the Hamiltonian is a constant of motion and the expression for the conjugate momentum $ p=e^{2F(x)} \dot{x}$  suggests that $M=e^{2F(x)}$ serves as a position dependent mass term. In fact equations with quadratic velocity dependance of the type considered here naturally arise in the Newtonian formulation of the equation of motion of a particle with a variable mass.
Clearly then, the trajectories for arbitrary intial condition $(x_0, y_0)$ where $ y=\dot{x}$ are given by
\be\label{eq:E7} \frac{1}{2}e^{2F(x)} y^2+V(x)=\frac{1}{2}e^{2F(x)} y_0^2+V(x_0).\ee

\bigskip

In terms of the canonical momentum, $p=e^{2F(x)} \dot{x}$, the Hamiltonian $H$ becomes 
\be H = \frac{p^2}{2e^{2F(x)}} + V(x).
\ee
Defining a new set of canonical variables 
\be
P := \frac{p}{e^{F(x)}} \qquad \hbox{ and } \qquad Q = \int_{0}^{x}e^{F(s)}ds = \Psi(x),
\ee
the Hamiltonian takes the appearance 
\be H = \frac{1}{2}P^2 + V(\Psi^{-1}(Q)) = \frac{1}{2}P^2 + U(Q), \qquad \hbox{ where } \,\,\,\,\, U = V \circ \Psi^{-1}
\ee
and corresponds to that of a particle of unit mass provided $\Psi(x)$ is invertible. In the following we consider a simple example
in which $f(x) = \hbox{ constant} $ and $g(x)=x$. Let $f(x) = 1/2$, so that $F(x) = x/2$. Then momentum and coordinate becomes
$P = e^{x/2}y$ and $Q = 2e^{x/2}$ respectively, and $V(x) = e^{x}(x-1) + 1$. Thus in terms of new coordinates the Hamiltonian has the following form
\be H = \frac{1}{2}P^2 + \frac{Q^2}{4}\ln (\frac{Q^2}{4} - 1). \ee

\begin{figure}[H]
\hfill{}\subfloat[$f(x)=\frac{1}{2}$, $g(x) = x$]{

\includegraphics[width=6cm,height=6cm,keepaspectratio]{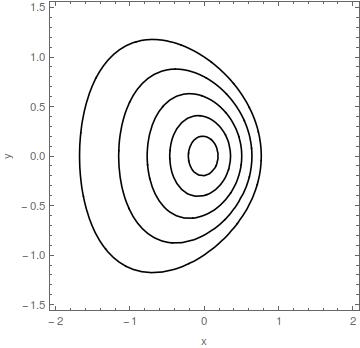}
}
\subfloat[$f(x)=3x$, $g(x) = x + x^3$]{

\includegraphics[width=6cm,height=6cm,keepaspectratio]{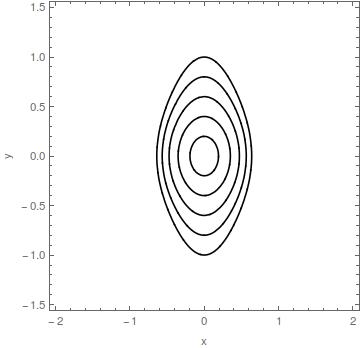}

}\hfill{}

\caption{Phase space orbits.}
\label{fig:q1}

\end{figure}

Figure [\ref{fig:q1}] shows some of the trajectories for the Hamiltonian in equation (\ref{eq:E7}) with different initial
condition and it is clear that the origin $(0, 0)$ is a centre.

\section{Quadratic damping}

It is plain that (\ref{E1}) can not describe a system with a quadratic damping as the  term involving $\dot{x}^2$ does not change sign and oppose the motion when the velocity reverses its sign. To remedy this feature it is necessary to split (\ref{E1}) into two parts  and write
\be\label{f1} \ddot{x}+f(x)\dot{x}^2 +g(x)=0,\;\;\;\dot{x}>0,\ee
\be\label{f2} \ddot{x}-f(x)\dot{x}^2 +g(x)=0,\;\;\;\dot{x}<0.\ee
Let us denote the Hamiltonians associated with the pieces by
\be\label{f3} H^{\pm}=\frac{1}{2}e^{\pm 2F(x)} y^2+V^{\pm}(x)\ee
with the superscript $\pm$ standing for $\dot{x}=y>(<)0$. Furthermore it will be assumed that the initial point $(x_0, y_0)$ with $(y_0>0)$ to be such that $V^+(x_0)=0$ and $F(x_0)=0$.
Thus when motion commences from the initial point then the trajectory is defined by $H^+=K_0^+=y_0^2/2$ or in explicit form
$$\frac{1}{2}e^{+ 2F(x)} y^2+V^{+}(x)=\frac{1}{2}y_0^2.$$
This trajectory crosses the $x$-axis first at say $x=x_1$ when the velocity $y=0$. Consequently the point of intersection $x_1(>0)$ which denotes the maximum amplitude is determined from the equation
$V^+(x_1)=y_0^2/2$, i.e., from
\be \label{E7}\int_{x_0}^{x_1}e^{2F(s)} g(s) ds =\frac{1}{2}y_0^2.\ee
Continuing the trajectory to below the $x$-axis means that it is now determined by the equation
$$H^-=\frac{1}{2}e^{-2F(x)} y^2+V^-(x)=K_1^-$$ where the constant $K_1^-$ is determined by the new initial condition $(x_1, 0)$ which represents the first turning point. This gives $K_1^-=V^-(x_1)$ so that beneath the $x$-axis the trajectory is given by
$$\frac{1}{2}e^{-2F(x)} y^2+V^-(x)=V^-(x_1).$$
At the next turning point we have $(x=x_2, y=0)$ and \be V^-(x_2)=V^-(x_1),\;\;\;(x_2<0)\ee whcih allows for the determination of $x_2$. After this we are again above the $x$-axis and the trajectory is given by $H^+=K_2^+$ with $K_2^+=V^+(x_2)$. Continuing in this manner we may summarise the points of intersections with the $x$-axis by stating that:\\
\noindent if $i=$odd then $x_{i+1}$ is determined by $$V^-(x_{i+1})=V^-(x_i)$$ and  if $i=$even then
$x_{i+1}$ is determined by $$V^+(x_{i+1})=V^+(x_i).$$

\subsection{Closed and Damped Orbits}

\begin{figure}[H]
\hfill{}\subfloat[$f(x)= 0.5$]{

\includegraphics[width=6cm,height=6cm,keepaspectratio]{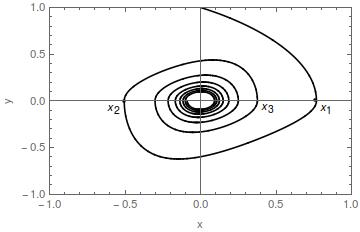}
\label{fig:p1a}
}
\subfloat[$f(x)=3x$]{

\includegraphics[width=6cm,height=6cm,keepaspectratio]{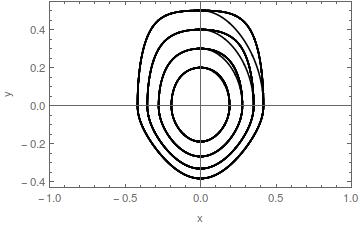}
\label{fig:p1b}
}\hfill{}

\caption{Phase space orbits.}
\label{fig:p1}

\end{figure}

Systems given by (\ref{f1}) and (\ref{f2}) do not always show damped behaviour. The qualitative features of such systems depends on $f(x)$. To study such systems, we consider two cases: $f(x)= 0.5,\,and\,f(x)=3x$ with $g(x)=x,\,and\,g(x)=x+x^{3}$,
respectively. Figure [\ref{fig:p1}] shows the plot for both the cases in phase
space. When $f(x) = 0.5$ the amplitude continuously diminishes and the trajectory is a spiral. Table \ref{table:t1} below 
gives the values of the amplitudes for each half of the cycle. In this case the dampening force has the same sign in each of the four quadrants as the linear damped oscillator.\\

\begin{table}
\hfil{}\begin{tabular}{|cc|ccc}
\cline{1-2} 
 & Amplitudes ($|x_{n}|$) &  &  & \tabularnewline
\cline{1-2} 
$x_{1}$ & 0.7680390470134656 &  &  & \tabularnewline
$x_{2}$ & 0.5049710640693359 &  &  & \tabularnewline
$x_{3}$ & 0.37698121404812546 &  &  & \tabularnewline
$x_{4}$ & 0.3009607508865283  &  &  & \tabularnewline
$x_{5}$ & 0.2505262046138416  &  &  & \tabularnewline
$x_{6}$ & 0.21459868739856763 &  &  & \tabularnewline
$x_{7}$ & 0.18769746691321143 &  &  & \tabularnewline
$x_{8}$ & 0.166796834446442  &  &  & \tabularnewline
$x_{9}$ & 0.15008870326261628 &  &  & \tabularnewline
\cline{1-2} 
\end{tabular}\hfill{}\caption{Table of amplitudes of each half of the cycle.}
\label{table:t1}
\end{table}

 However when $f(x) = 3x$ one finds that the orbits are
closed. Figure [\ref{fig:p1}\protect\subref{fig:p1b}] shows the orbits for four initial conditions. In this case the dampening force alternates in sign in each of the four quadrants.

\begin{figure}
\hfill{}$f(x)=\frac{1}{2}$\hfill{}

\hfill{}\subfloat[$V^{+}(x)$]{\includegraphics[width=5cm,height=5cm,keepaspectratio]{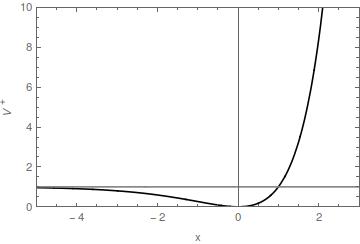}

}\subfloat[$V^{-}(x)$]{

\includegraphics[width=5cm,height=5cm,keepaspectratio]{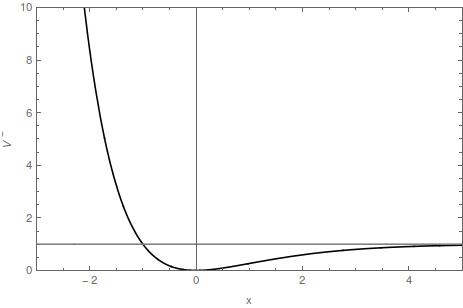}

}\hfill{}

\hfill{}$f(x)=3x$\hfill{}

\hfill{}\subfloat[$V^{+}(x)$]{

\includegraphics[width=5cm,height=5cm,keepaspectratio]{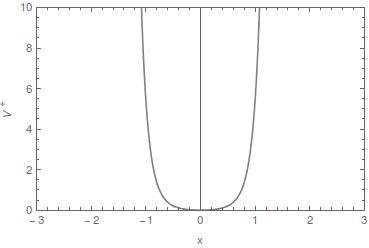}

}\subfloat[$V^{-}(x)$]{

\includegraphics[width=5cm,height=5cm,keepaspectratio]{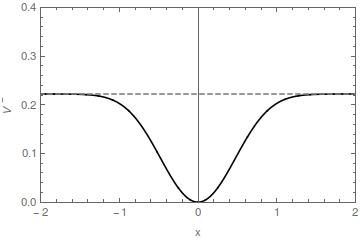}

}\hfill{}

\caption{Potential plots.}
\label{fig:p2}

\end{figure}

In our case, the potentials are denoted by $V^{\pm}(x)$. Figure [\ref{fig:p2}] shows
the potential plots for both the cases under study. From the figure,
it is clear that there are initial conditions for which bounded solutions exists in both the cases.

\begin{figure}[H]

\hfill{}\subfloat[$f(x)=\frac{1}{2}$]{

\includegraphics[width=6cm,height=6cm,keepaspectratio]{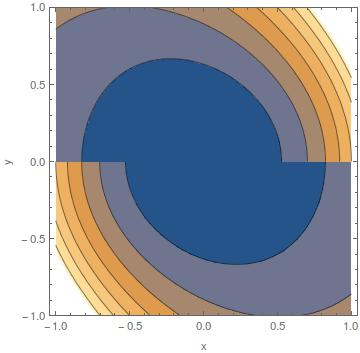}

}\subfloat[$f(x)=3x$]{

\includegraphics[width=6cm,height=6cm,keepaspectratio]{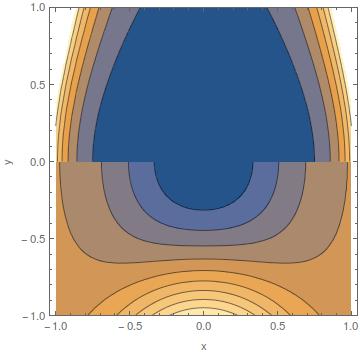}

}\hfill{}

\caption{Hamiltonian plots.}
\label{fig:p3}

\end{figure}

Now, consider the kinetic part. In our case the mass term is position
dependent and plays the key role in whether there is closed or damp
orbit. The main idea is the mass term depends on $F(x)$ only. Consider
 figure [\ref{fig:p3}] of contour plots of Hamiltonian in the two cases. In
the first case, $f(x)$ is constant and so even and therefore $F(x)$
is odd. While in the second case, $f(x)$ is odd function and $F(x)$
is even and its evident from the plots that Hamiltonian is symmetric
about the $y-axis$.

\begin{figure}[H]

\hfill{}\subfloat[$f(x)=x^{2},\,g(x)=x^{5}$]{

\includegraphics[width=6cm,height=6cm,keepaspectratio]{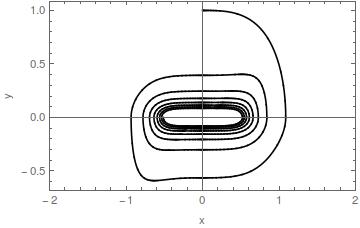}

}\subfloat[$f(x)=x^{3},\,g(x)=x^{7}$]{

\includegraphics[width=6cm,height=6cm,keepaspectratio]{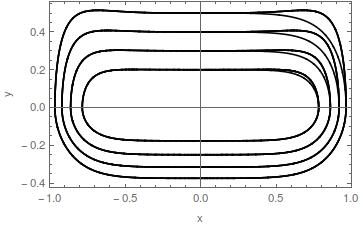}

}\hfill{}

\caption{Phase space orbits for higher degree functions.}
\label{fig:p4}

\end{figure}

Figure [\ref{fig:p4}] depicts the behaviour of higher degree functions.

\subsection{Analytic results based on Cheillini integrability condition}
It is interesting to observe that invocation of the Cheillini condition for integrability allows us to determine analytically the maximum amplitudes. This may be accomplished by noting that as
$$ V^{\pm}(x_i)=\int_{x_0}^{x_i}e^{\pm 2 F(s)} g(s) ds \;\;\;\mbox{with}\;\;\;F(x)=\int_{x_0}^x f(s) ds\Longleftrightarrow F^\prime(x) =f(x)$$ so integrating by parts we obtain
$$ V^{\pm}(x_i)=\frac{1}{(\pm 2)}e^{\pm 2F(s)}\left(\frac{g(s)}{f(s)}-\frac{\mu}{(\pm 2)}\right)\Big|_{x_0}^{x_i}$$ where the Cheillini integrability condition
\be \frac{d}{dx}\left(\frac{g}{f}\right)=\mu f(x), \;\;\;\mu=const. (\ne 0),\ee
has been used. From this condition we have
$\frac{g}{f}=\mu F(x) +\nu$  where $\nu$ is an arbitrary constant of integration.  Of course, if $f$ and $g$ are given beforehand then the values of the constants $\mu$ and $\nu$ can be simply read off from the above ratio assuming the Cheillini condition to hold. By
defining $\eta_{\pm}=(\pm2\nu-\mu)/\mu$ and the function $G_\pm(x)=\pm2F(x)+\eta_\pm$ the last expression for the potential can be put into the form
\be\label{E10}V^{\pm}(x_i)=\frac{\mu}{4}e^{-\eta_{\pm}}\left[e^{G_\pm(x_i)}G_\pm(x_i)-e^{G_\pm(x_0)}G_\pm(x_0)\right].\ee
 As mentioned in (\ref{E4}) the first turning point is given by a solution of $V^+(x_1)=y_0^2/2$ which is positive or more explicitly by a positive solution of 
 \be e^{G_+(x_1)}G_+(x_1)=e^{G_+(x_0)}G_+(x_0)+\frac{2y_0^2}{\mu}e^{\eta_+}.\ee In general it is possible to find recursively the subsequent turning points  which are the solutions with proper signs of the equation
 \be\label{E12} e^{G_\pm(x_{i+1})}G_\pm(x_{i+1})= e^{G_\pm(x_{i})}G_\pm(x_{i}), \;\;\;+(-)\;\;\mbox{for}\;\;i=\mbox{even(odd)}.\ee
Eqn.(\ref{E12}) has the appearance $e^XX=Y$ which is the the standard form of the Lambert equation and its solutions are given by the Lambert $W$ function, i.e, $X=W(Y)$. Hence it follows that
\be G_{\pm}(x_{i+1})=W\left(e^{G_\pm(x_{i})}G_\pm(x_{i})\right),\ee from which we can determine formally 
\be x_{i+1}=G_\pm^{-1}\left(W\left(e^{G_\pm(x_{i})}G_\pm(x_{i})\right)\right),
\;\;\;+(-)\;\;\mbox{for}\;\;i=\mbox{even(odd)}\ee if the function $G_\pm$ is invertible. We therefore conclude that if the Cheillini condition is fulfilled by the functions $f$ and $g$ then the maximum amplitudes can be determined recursively in terms of the Lambert $W$ function.

\subsection{Analysis of the period function}
Starting from the initial point $(x_0, y_0)$ if the first intersection with the $x$-axis is at $x=x_1 (>0)$ then subsequently the trajectory is determined by the condition 
$$\frac{1}{2}e^{-2F(x)}y^2+V^-(x)=V^-(x_1).$$ 
If the next turning point, i.e., the next intersection with the negative $x$-axis is at $x=x_2$ then the time for the transit of the half-cycle from $x_1$ to $x_2$ is given by 
\be\tau_{12}=\frac{1}{\sqrt{2}}\int_{x_2}^{x_1}\frac{e^{-F(x)} dx}{\sqrt{V^-(x_1)-V^-(x)}}.\ee Similarly the time taken for the transit of the next-half cycle from $x_2$ on the negative $x$-axis to $x_3$ on the positive $x$-axis is given by 
\be \tau_{23}=\frac{1}{\sqrt{2}}\int_{x_2}^{x_3}\frac{e^{F(x)} dx}{\sqrt{V^+(x_2)-V^+(x)}}.\ee
Thus the time taken for the completion of the first cycle is $T_1=\tau_{12}+\tau_{23}$, viz
\be\label{E15} T_1=\frac{1}{\sqrt{2}}\left[-\int_{x_1}^{x_2}\frac{e^{-F(x)} dx}{\sqrt{V^-(x_1)-V^-(x)}}
+\int_{x_2}^{x_3}\frac{e^{F(x)} dx}{\sqrt{V^+(x_2)-V^+(x)}}\right].\ee
In fact it is straightforward to  generalize this formula for the $n$-th cycle which is given by 
\be\label{E16} T_n=\frac{1}{\sqrt{2}}\left[-\int_{x_{2n-1}}^{x_{2n}}\frac{e^{-F(x)} dx}{\sqrt{V^-(x_{2n-1})-V^-(x)}}
+\int_{x_{2n}}^{x_{2n+1}}\frac{e^{F(x)} dx}{\sqrt{V^+(x_{2n})-V^+(x)}}\right].\ee

The energy dissipated in the $n$-th cycle is given by 
\be \Delta E_n=V^-(x_{2n-1})-V^+(x_{2n}).\ee
Table \ref{table:t2} gives the values of time periods and change in energy for the case $f(x)=0.5$ and Table \ref{table:t3} gives the time periods for the closed orbits in the case $f(x)=3x$ for different initial conditions. In our case we have assumed that the energy is given by the Hamiltonian and differs from the expression used in \cite{Cveticanin1}.

\begin{table}
\hfil{}\begin{tabular}{|c||cc|cc|c|}
\cline{1-3} 
Cycle($n^{th}$) & $T_{n}$ & $\Delta E_{n}$  & \tabularnewline
\cline{1-3} 
$1$ & 6.360502498547287 & 0.08805266518005395   & \tabularnewline
$2$ & 6.308066845017076 & 0.01833863366586541   & \tabularnewline
$3$ & 6.295534499523865 & 0.006618943128696708  & \tabularnewline
$4$ & 6.29057350609154  & 0.0031022724220284292 & \tabularnewline
$5$ & 6.288104015677968 & 0.0016959179654737477 & \tabularnewline
$6$ & 6.286695917618085 & 0.0010263408918240735 & \tabularnewline
$7$ & 6.285816631877343 & 0.0006676772782294726 & \tabularnewline
$8$ & 6.285230961390923 & 0.0004584700142998832 & \tabularnewline
$9$ & 6.284821348882355 & 0.0003283012470082225 & \tabularnewline
\cline{1-3} 
\end{tabular}\hfill{}\caption{Time Periods and Energy change in each cycle for $f(x)=0.5$.}
\label{table:t2}
\end{table}

\begin{table}
\hfil{}\begin{tabular}{|c||c|cc|}
\cline{1-2} 
i=(${x_{0}^{i},y_{0}^{i}}$) & $T^{(i)}$ & \tabularnewline
\cline{1-2} 
${(0,0.2)}$ & 6.197928744154146  & \tabularnewline
${(0,0.3)}$ & 6.111126244536706   & \tabularnewline
${(0,0.4)}$ & 6.014565567199962  & \tabularnewline
${(0,0.5)}$ & 5.918345639978987  & \tabularnewline
\cline{1-2} 
\end{tabular}\hfill{}\caption{Time Periods of Cycles for $f(x)=3x$ for different initial conditions.}
\label{table:t3}
\end{table}

\bigskip

\section{Conclusion}
In this paper,  a Li\'enard type equation with quadratic damping has
been considered. We have shown  that by imposing the Cheillini condition of integrability on the functions $f$ and $g$ one 
can demonstrate many of the previous examples into a compact scheme. Incidentally the Cheillini condition is typically encountered in the 
context of integrability of the standard Li\'{e}nard equation in course of its transformation to the first-order Abel equation of the first kind and also while finding a Lagrangian/Hamiltonian description of the Li\'{e}nard equation. 
Our next target is to study generalized oscillatory equation with a non-negative real-power
restoring force and quadratic damping and seek for the solution and other dynamical features of this type of equations.
In particular, we wish to generalize Kovacic et al. \cite{KR} model where they have considered purely 
non-linear restoring force of the type $F(x) = k sgn(x)|x|^{\alpha}$ ,

\section*{Acknowledgement}

The authors are grateful to Professor Livija Cveticanin for various
useful discussions
They are also thankful to Professor Asok Mallik for his 
interest and encouragement  and one of us (PG) wishes to acknowledge ``Dynamics of Complex Systems, DCS-2016'' programme at ICTS-TIFR
for their gracious hospitality where part of the work has been done.

\end{document}